\documentstyle[aps,preprint]{revtex} 
\begin{document}
\draft

\title{ Born Effective Charge Reversal and Metallic Threshold State 
at a Band Insulator-Mott Insulator Transition}  

\author{ N. Gidopoulos$^{1,2}$, S. Sorella$^{1,2}$, 
and E. Tosatti$^{1,2, 3}$} 
\address{$^1$ International School for Advanced Studies, (SISSA) 
Via Beirut 2-4, I-34014 Trieste, Italy\\
$^2$  Istituto Nazionale di Fisica della Materia (INFM), Unita' Trieste
SISSA, and \\
$^3$ International Centre for Theoretical Physics, P.O. Box 586, I-34014
Trieste, Italy 
} 

\date{\today} 
\maketitle
\vskip -0.75 truecm
\begin{abstract}
\widetext {}

We study the quantum phase transition between a band (``ionic'') 
insulator and a Mott-Hubbard insulator, realized at a critical value 
$U=U_c$ in a  bipartite Hubbard model with two inequivalent sites,
whose on-site energies differ by an offset $\Delta$. 
The study is carried out both in $D=1$ and $D=2$ (square and honeycomb 
lattices), using exact Lanczos diagonalization, finite-size scaling, and
Berry's phase calculations of the polarization. The Born effective 
charge jump from positive infinity to negative infinity
previously discovered in $D=1$ by Resta and Sorella 
is confirmed to be directly connected with the transition from the band
insulator to the Mott insulating state, in agreement 
with recent work of Ortiz {\it et al.}. In addition, symmetry is  
analysed, and the transition is found to be associated with a 
reversal of inversion symmetry in the ground state, of 
magnetic origin. We also study the $D=1$ excitation spectrum 
by Lanczos diagonalization and finite-size scaling. Not only
the spin gap closes at the transition, 
consistent with the magnetic nature of the Mott state, but also
the {\em charge} gap closes, so that the intermediate state
between the two insulators appears to be metallic.
This finding, rationalized within unrestricted 
Hartree-Fock as due to a sign change of the effective on-site energy
offset  $\Delta$ for the minority spin electrons, underlines the 
profound difference between the two 
insulators. The band-to-Mott insulator transition is also studied
and found in the same model in $D=2$.  There too we find an associated, 
although weaker, 
polarization anomaly,
with some differences between square and honeycomb lattices. 
The honeycomb lattice, which does not possess an inversion symmetry, 
is used to demonstrate the possibility of an inverted piezoelectric
effect in this kind of ionic Mott insulator.  

\end{abstract}

\widetext
\section{Introduction}

Standard discussions of the Mott-Hubbard transition are generally concerned
with lattices of equivalent sites. At zero temperature, the metal-insulator
transition develops when the on-site electron-electron repulsion $U$ reaches
some critical value $U_c$ which, usually, also corresponds to the onset 
of characteristic magnetic correlations.
In this paper we are concerned with the less common case where 
the system is ionic, encompassing ``anions'' and ``cations''.
Earlier workers, including Nagaosa and Takimoto\cite{nagaosa}, and Egami,
Ishihara and Tachiki \cite{egami} considered
the simplest two-site ionic generalization of the Hubbard model (henceforth
dubbed Ionic Hubbard Model, IHM), which exhibits 
the transition from a band insulator to a Mott insulator.
Such a system is, in the absence of electron-electron repulsions, 
$U$ = 0, a regular band insulator . 
As $U$ increases above some critical value $U_c$,
a band insulator - Mott-Hubbard insulator
transition is expected to take place. When $U$ is sufficiently
large, the inequivalence between anion and cation should in fact become
irrelevant, and the ground state of a large $U$ system of equivalent
sites is Mott-insulating, with antiferromagnetic correlations.
However, owing to the residual 
inequivalence of the two ionic sites, it will also exhibit other 
properties, which have only partly
been explored so far. Of special interest is the anomalous behaviour 
of the polarization of the ionic solid across the transition. A very
interesting quantity in this regard is the Born effective charge $Z^*$
associated with an infinitesimal ``dimerizing'' displacement of the 
ionic lattice, corresponding to a frozen  $q=0$ optical phonon.
Resta and Sorella \cite{sorella} studied the IHM in $D=1$ using
Lanczos diagonalization and found, by the Berry phase method, very
striking indications of a polarization anomaly at  $U=U_c$. Ortiz, Ordej\'on, 
Martin and Chiappe \cite{ortiz2} proposed a simple Hartree-Fock
explanation to the anomaly, namely that at a first order magnetic
transition between an AF insulator and a band magnetic insulator,
the polarization may also abruptly jump.

A number of important questions are apparently 
 still open at this stage, in particular
\begin{enumerate}
\item what is the physical origin of the polarization
anomaly at the band-Mott transition, and what is its connection with symmetry?
\item what is the nature precisely at $U = U_c$,  of the ``threshold''
state between the band and the Mott insulator?
\item how does the polarization anomaly depend on the dimensionality? 
In particular, will
it survive in 2D and 3D, and if so, with what strength?
\item bearing in mind that most experiments measure just only ${Z^*}^2$,  
can we identify a simple experimentally accessible quantity which could 
signal the $Z^*$ anomaly in magnitude and sign?
\end{enumerate}

In this paper we set out  to discuss principally these questions. 
We will do it by studying more closely the same simple ionic 
Hubbard model \cite{egami,sorella,ortiz2}, in particular by analysing
the symmetry of its possible ground states, by calculating effective charges
and excitation spectra, and generally by seeking to understand 
its properties by exact diagonalizations supplemented when necessary
by the simple Hartree-Fock approximation.

Firstly, we find that the Mott insulator does, for appropriate
boundary conditions, possess odd
symmetry under inversion, contrary to the  ionic band insulator, which
has even parity.  Secondly, Hartree-Fock reveals 
that one clue to the polarization anomaly
lies in an effective reversal of the on-site energy offset 
(accompanied with the vanishing of the associated band gap), taking place 
{\em for the minority spin species only} at $U_c$. Thirdly, in agreement
with Hartree-Fock, exact results suggest that at 
$U_c$ not only the spin gap, but also the true charge gap vanishes , 
indicating a {\em metallic} ``threshold'' state poised precisely at the 
brink between the band insulator and the Mott insulator. Fourthly, 
a fresh study of two different 2D lattices, 
namely square and honeycomb, shows that 
a direct band insulator-Mott Hubbard
insulator transition and the associated polarization anomaly may
survive in higher
dimensions too, particularly in the 2D honeycomb lattice. Here,
the anomaly is weaker than in D=1 ( a jump instead of 
a divergence). Finally, we propose the piezoelectric effect in 
a non-centrosymmetric 
lattice, here exemplified precisely by the 2D honeycomb lattice, as the
experimentally accessible quantity that will directly and strikingly 
change sign at the band
insulator-Mott insulator transition.

The polarization calculations are 
carried out using the Berry phase technique, first introduced 
by King-Smith and Vanderbilt \cite{king}
and by Resta \cite{resta}, and further extended to a general many-body case 
by Ortiz and Martin \cite{ortiz1}, and applied to the IHM by Resta 
and Sorella. 

The rest of this paper is structured as follows. In section \ref{sec:model} we 
introduce the ionic Hubbard model and describe briefly
the classical limit, $t=0$, useful for understanding later
the full quantum case. In section \ref{sec:magnetism}
we first present a discussion of the magnetic behavior expected by 
the model, followed by a more detailed discussion of symmetry, and by
a finite-size study
of the level crossings connected with the transition at $U_c$.
Section \ref{sec:hf} contains the Hartree-Fock theory of the band- 
Mott insulator transition, and of the polarization anomaly. Section
\ref{sec:manybody} presents the full many-body calculation of the polarization,
done by means of Lanczos diagonalization plus Berry phase, 
here specialized  to the 2D honeycomb case, and finally in Section 
\ref{sec:discussion} we give a discussion of the possible detectability
of the transition via the piezoelectric effect, followed by our general 
conclusions.

\section{Ionic Hubbard model} \label{sec:model}

We consider the Hubbard hamiltonian in 1D (linear chain) and 2D bipartite
lattices (square and honeycomb lattices). All lattices being 
bipartite, they are composed of $A$ and $B$ sublattices. To simulate
ionicity, $A$ and $B$ are made inequivalent by onsite energies 
$\frac{\Delta}{2}$ and $- \frac{\Delta}{2}$ respectively. Because of the
energy difference $\Delta$ between the $A$ and the $B$ sublattices, 
the electron population of the $A$
sublattice is less than that of the $B$ sublattice for $\Delta > 0$ (or
viceversa for $\Delta < 0$).
The sublattices are connected by electron hopping . 
We assume a filling of one electron per site with
equal numbers of spin up and of spin down electrons.
Sites of the $A$ sublattice are denoted by $R_{A}$, those of the $B$ sublattice by
$R_{B} = R_{A} + \xi_{\mu}$, where the vectors $\xi_{\mu}$ 
connect an $A$ site with its neighbouring $B$ sites, $\mu=1,2,\ldots,\nu$.
In the linear chain, where the length of the unit cell is 2, $\mu$ = 0,1, $\xi_{0,1}$ = $\pm 1$. 
In the square lattice (square side of unit length),
$\mu$ = 0,1,2,3 and $\xi_{0,2}$ = $\pm$(1,0), $\xi_{1,3}$ = $\pm$(0,1). 
In the honeycomb lattice (unit length is the side of the hexagon),  $\mu$ = 0,1,2 and
$\xi_{0}$ = $(- {1 \over 2},{ \sqrt{3} \over 2})$, $\xi_{1}$ = 
$(1,0)$, $\xi_{2}$ = $(- {1 \over 2},- { \sqrt{3} \over 2})$.\\
The Hamiltonian is
\begin{equation}\label{hubbard}
H = - \sum_{ R_{A}, \mu , \sigma  } t_\mu  c^{\dag}_{ R_A + \xi_\mu,\sigma  } c_{ R_A ,\sigma } + {\rm h.c.}
+ U \sum_{R} n_{R \uparrow } n_{R \downarrow } + {\Delta \over 2} \left[
\sum_{R_A,\sigma }  c^{\dag}_{ R_A,\sigma } c_{ R_A,\sigma }
- \sum_{R_B,\sigma }  c^{\dag}_{ R_B,\sigma } c_{ R_B,\sigma } \right]
\end{equation}
where we have used standard notation, and U is the Hubbard onsite
electron-electron repulsive interaction.
Electrons hop with matrix elements $-t_\mu (t_\mu >0)$ 
between neighbouring sites from the $A$ to the $B$ sublattice 
along the $\xi_\mu$ directions.
We denote the ground state energy of our $N$-electron system as $E(N)$. 
When in need to distinguish 
between the number of spin up and spin down electrons, the ground state energy 
will be denoted by $E(N\uparrow, N\downarrow)$. 
Unless otherwise specified $N\uparrow = N\downarrow = N/2$.
We shall study the charge gap and the spin gap of the system defined respectively as: 
$\Delta E_{charge} = E(N+1)+E(N-1)-2E(N)$ and 
$\Delta E_{spin} = E(N\uparrow +1, N\downarrow -1) - E(N\uparrow, N\downarrow)$.\\

\subsection{Behavior in the classical limit, $t=0$} \label{sec:t0}

It is instructive at the outset to consider what happens when all $t_\mu=0$.
As sketched in Fig(\ref{class}) if we set the hopping $t_\mu = 0$ in the 
Hamiltonian (\ref{hubbard}) 
the model is classical. It has a first order transition 
at $U = \Delta$ with a change of the macroscopic polarization per unit cell 
$\Delta P = e  a/2$. Both the charge and spin gaps are finite and 
coincide in the region $U < \Delta$ where they are given 
by $\Delta E = \Delta - U$. For $U > \Delta$ 
the charge gap remains finite $\Delta E_{charge} = U - \Delta$ while the spin
gap vanishes. Note that precisely at $U = \Delta$, where the transition
takes place, both 
the charge and the spin gap
vanish. This kind of transition s expected to  persist for $t>0$, where it
takes a standard band insulator,  with a charge and a spin gap, 
over to an antiferromagnetic insulator, with a charge gap, and 
gapless spin excitations for large $U$. 

\section{Magnetism, Symmetry, Level Crossing and Polarization Jump} \label{sec:magnetism}

In this section we discuss the polarization properties of an electron
system close to an antiferromagnetic 
transition. As discussed by Ortiz and Martin antiferromagnetism appears
to play a crucial role and remarkably affects the behaviour of the
polarization.\cite{ortiz2}

Let us consider a finite electron hopping $t_\mu$.
For $U=0$, one electron per site, the model is in fact 
described by  a completely filled, spin independent band, separated by
a finite gap ($\sim \Delta$, for large $\Delta$) from the second, empty band. 
In the opposite limit of large  $U$ the standard strong coupling 
analysis of the Hubbard model leads to a 
charge gap $\sim U$, and to the 
well known Heisenberg Hamiltonian \cite{anderson1} $H_J$ for spins:
\begin{equation}\label{heis}
H_J = J \sum_{<i,j>} \vec S_i \cdot \vec S_j  
\end{equation}
with an antiferromagnetic superexchange coupling $J ={ 4 t^2\over U} $. 
The mapping to the Heisenberg model implies that for large $U$ the
model has gapless spin excitations.  
In one dimension they have been named ``spinons''\cite{anderson1}, 
and can be derived   from the exact
Bethe ansatz solution of the 1D Heisenberg model.\cite{faddeev} 
In higher dimensions, where antiferromagnetic long range order
is believed to exist at $T =0$ in 2D \cite{reger,runge,chn}
and 3D\cite{lieb}, the gapless modes are spin waves. 

A band insulator - Mott insulator transition should therefore occur  
at some finite coupling  $U_c$. Of course, $U_c$ will differ from 
its classical value $U_c^0=\Delta$. There will be in general the
possibility of intermediate metallic phases covering a range of
$U$ values. Even if the insulator-insulator transition is direct,
quantum fluctuations may drive its character, for example from first 
order to second order, or else may split it into more than one 
transition \cite{fabrizio}. 
Quantum antiferromagnets with long range order for $U>U_c$ (or quasi long range
order in one dimension) can be described at the critical level, 
by the well known non linear sigma model, with action
\begin{equation} \label{sigma}
$$ S = {1\over 2 g} \int dx^{d+1}  (\vec \nabla \vec n)^2 $$
\end{equation}
where $\vec n$ is a unit vector describing the local orientation of the 
antiferromagnetic order parameter, and  
the  coupling constant $g$ depends on  the microscopic parameters 
 $U,\Delta$ 
of the model.
This effective model is well known to have a second order
transition predicting that a spin gap opens up continuously for $g> g_c$, 
alias $U< U_c$ \cite{chn}.

We will for simplicity  assume in the following that the transition is
unique and is always second order 
even though a Hartree-Fock calculation \cite{ortiz2} in the one
dimensional model and our own in the 2D square lattice indicates
the opposite. In fact, the Hartree-Fock approximation may fail anyway to
describe correctly the  order of the transition, as it is not appropriate to 
describe the gapless spin-wave excitations of the  model in the ordered phase. 
Contrary to the linear chain and the square lattice, 
the Hartree-Fock method however correctly predicts (see Sec. \ref{sec:hf}) 
a second order phase transition in 
the honeycomb lattice, at least for the parameter values studied  here.

The magnetic Mott insulator for  $U> U_c$ has a charge gap,
and is therefore fully described by the nonlinear sigma model,
at least in more than one dimensions.
This model does not present any further phase transition but 
the one at the critical coupling $g_c$, which is thus related to $U_c$. 
It is possible however that the charge transition to a band insulator 
might occur for $U$ values  different from $U_c$, as suggested
by Fabrizio {\em et al} \cite{fabrizio}. We shall return to this point below.

\subsection{D=1: symmetry, level crossing and critical $U_c$ }

In a finite system with -say- $L$ sites a level crossing  may occur 
 for some particular boundary conditions, if allowed by symmetry. 
In particular, in the one dimensional 
model, it can be easily proved that there exists  a finite value $U_c(L)$ 
at which  the ground state undergoes a change in the eigenvalue 
of the inversion symmetry operator $R$.
Inversion symmetry  around the  site  $i=0$    is  defined 
by the following relations:
\begin{eqnarray}
R c^{\dag}_{\displaystyle i,\sigma } R^{\dag} &=&  c^{\dag}_{\displaystyle
L-1-i,\sigma} ~~~{\rm for} ~ i=0,1,\cdots L-1 \label{inversion1} \\ 
 R |0>&=&|0> \label{inversion}
\end{eqnarray}
where $|0>$ is the electron vacuum state, by definition invariant under
inversion symmetry. 

The additional relation (\ref{inversion}) is necessary 
to completely  define the  inversion
transformation $\hat R$ in the whole Hilbert space.
Inversion does not interchange the $A$ and the $B$ sublattice
and clearly commutes with the hamiltonian (\ref{hubbard})  for any $U$,
provided the boundary conditions are real , namely periodic or antiperiodic 
(see Sec. \ref{sec:manybody}).

Since $R^2 =I$, the identity, the inversion has obviously
two eigenvalues, $\pm 1$. We will show in  the following that 
the ground state $|\psi_0>_U$ satisfies  
$ R |\psi_0>_U= |\psi_0>_U$ for $U=0$, whereas  
 for large $U$ there is a change of sign and 
 $ R |\psi_0>_U= - |\psi_0>_U$, so that {\em a level  
crossing must occur at an intermediate coupling } $U=U_c(L)$.

In the non-interacting system the ground state is a direct 
product of a spin up and a spin down Slater determinants, both possessing
 the same orbitals of the lowest band. Both 
Slater determinants have  a definite parity $R_\sigma=\pm 1$  and the 
inversion eigenvalue  of the global wavefunction is given by their product 
$$R=R_\uparrow \times R_\downarrow =1$$. Hence the band insulating 
 state is even under inversion.
 
In the large-$U$ Mott insulator instead we can use the mapping to 
the Heisenberg hamiltonian 
(\ref{heis}) , whose ground state 
in arbitrary dimensions can be generally written as \cite{lieb}:
$$ |\psi_0>_{U\to \infty} = \sum_{i_1,i_2,\cdots i_{L/2} } 
\Phi (i_1,i_2,\cdots i_{L/2}  )
S^-_{i_1} S^-_{i_2} \cdots S^-_{i_{L/2}} |F> $$
where $S^-_i= c^{\dag}_{\downarrow i} c_{\uparrow i}$ is the spin lowering 
operator at the site $i$,  
$|F>=\prod\limits_i c^{\dag}_{\uparrow,i} |0>$  is the ferromagnetic state 
along the spin-up direction, and  
 the wavefunction $\Phi$ is real. $\Phi$ is also subject to the well  
known (``Marshall sign'') condition \cite{lieb}, i.e. 
the sign of the wavefunction is determined by the number of 
spin flips in the $B$ sublattice ($i$ odd): 
$$ \Phi (i_1,i_2,\cdots,i_{L/2}) (-1)^{\displaystyle \sum_{k=1}^{L/2} i_k }  > 0.$$ 
 The action of the  inversion symmetry $R$ 
on the spin lowering operators can be easily found by applying the 
definition given in   Eqs.(\ref{inversion},\ref{inversion1}),
 namely 
$R S^-_{\displaystyle i} R^{\dag} = S^-_{\displaystyle L-1-i} $, 
and thus $R$  
maps an element of the basis
 $S^-_{i_1} S^-_{i_2} \cdots S^-_{i_{L/2}} |F>$ to another one 
$S^-_{i^\prime_1} S^-_{i^\prime_2} \cdots S^-_{^\prime i_{L/2}} R |F>$ 
where $i^\prime= L-1-i$  and $R |F>=\prod\limits_i c^{\dag}_{L-1-i} |0> = \pm 
|F>$. The overall sign $\pm$ in the latter equation  represents 
just the inversion eigenvalue of the ferromagnetic state $|F>$ and 
can be  determined  using the canonical
anticommutation rules to restore the order of the creation 
operators $c^{\dag}_i$  
 after the application  of the inversion operator 
$R$ to the ferromagnetic state $|F>$.
Since inversion symmetry does not 
change the Marshall sign we arrive to the conclusion that the inversion
eigenvalue of the Heisenberg wavefunction coincides with  the inversion 
eigenvalue of the ferromagnetic state $|F>$ which is simple to compute.

Doing that we find that for $U\to \infty$, the inversion eigenvalue can
change sign depending on the boundary conditions (b.c.):
 \begin{eqnarray} 
R&=&(-1)^{\displaystyle L/2 +1}     ~~~ {\rm for ~ periodic b.c. } \nonumber \\
R & =&  (-1)^{\displaystyle L/2 }   ~~   {\rm for ~ antiperiodic b.c.}
\end{eqnarray}

This finally proves our initial statement; in particular a level crossing 
from an even state to an odd one {\em has} to 
occur in a periodic ring with $L=4 n$ or in an antiperiodic one with $L=4 n+2$.
On the other hand, there will not necessarily be a level crossing 
in a periodic ring
with $L=4 n+2$ and an antiperiodic one with $L=4 n$. In summary, we
conclude that the demise of the band insulator occurs via a symmetry
change, whose finite-size signature is a parity switch from even to odd
in the appropriate boundary conditions.

\subsection{Consequences on the calculation of the
 polarization}\label{sec:chspgap}
As will be discussed in Sec.(\ref{sec:manybody}) 
the change of polarization in a
many-body system  can be obtained using a form of averaging over the boundary
conditions. Thus the averaging  necessarily include both
periodic (PBC) and antiperiodic (APBC) boundary conditions. Now, 
in one or in the other, depending on L,  a level crossing will necessarily
occur at some finite  $U_c(L)$. 
>From the theory of polarization a strong variation 
of the polarization can be expected as a function
of $U$  around $U_c(L)$ even in presence of a perturbations 
such as  a dimerization $\delta$ (see Sec.(\ref{sec:manybody})). 
Therefore , within the hypothesis that there exists only a well defined
Mott  transition at a critical value $U_c$ in the thermodynamic limit, we
may expect that:
\begin{equation}  
U_c(L) \to U_c  ~~~ {\rm for}~ L \to \infty
\end{equation}
that is, the $U$ value where the level crossing occurs for large size 
is just the critical value of the magnetic transition.

\subsection{Charge and Spin Gaps in D=1}

Understanding charge and spin excitation gaps is a crucial point.
We have studied these quantities in the D=1 case as a function of $U/t$, 
performing calculations on finite rings
with PBC or APBC.  
By considering the sequence of closed shells 
with one electron 
per site, there is no level crossing and a finite size scaling analysis 
can be safely applied to the charge and spin gaps (see the end of 
Sec.\ref{sec:model} for their definitions). 
For a general finite size system, the lowest order correction to
any gap should be of the form $A \over L^2$. We have used a three parameter
fit 
\begin{equation} \label{fit}
\Delta_L = \sqrt{\Delta^2 + {A \over L^2} + {B \over L^3} + \ldots}
\end{equation} 
including  also a higher order $L^{-3}$ term, to improve the accuracy.
In Fig. (\ref{spinchargecl}) we show the finite size  calculations of the gaps
for the 6 and the 12 site ring, as well as the result 
extrapolated to the thermodynamic limit 
with  the finite size scaling (\ref{fit}).  
In Fig. (\ref{spinchargeop}) 
we also present the results for the
spin and charge gaps for open shell rings of 6 and 12 sites. 
The overall behavior of the gaps  for the largest open shell ring (L=12) 
is in agreement  with the infinite size extrapolation 
of the closed shell rings,  supporting the validity of our finite size scaling.
Starting with the band insulator at small $U$, and increasing $U$,
we find that both charge and spin gaps are very close, and decrease
together until they appear to close at some  $U_c$.
For $U>U_c$, the charge gap turns sharply upwards, while the spin gap doesn't.
Precisely at $U_c\simeq 2.75 t$, our fit suggests finite size gap corrections 
of the form $1\over L$, which implies not only spin, but also charge gapless 
excitations. 
This being the case, the system at $U_c$ is metallic. The nature of this metal
is unknown and deserves further investigation.\cite{sholl}

>From our calculations  it is hard to say whether charge and spin gaps
will vanish  at exactly the same $U_c$, or at two slighly different 
values, as  very recently proposed in \cite{fabrizio}.
Nonetheless it is suggestive that large finite 
size charge gaps and small spin gaps become slightly inverted after
extrapolation, which goes precisely in the direction of a charge gap 
closing at a slightly smaller  $U$ than the spin gap. 

We defer all discussion of the possible two-transition scenario to the 
work of Fabrizio {\it et al.} \cite{fabrizio} and we will not further dwell
on it in this paper, where we consider for simplicity a single $U_c$.

\subsection{Extension to higher dimensions: D=2}
In the previous analysis of a level crossing in the 
model (\ref{hubbard}) we did not explicitly use the 
exact Bethe ansatz solution of 1D systems. 
In  fact the result 
that the inversion symmetry  $R$ for large $U$ has the same eigenvalue of the 
corresponding ferromagnetic state $|F>$ remains valid  also in D=2,
an so does the evenness of the band insulator at small $U$.

Unfortunately, unlike  D=1, the D=2 inversion symmetry, transforming 
$(x,y) \to ( L-1-x,L-1-y) $  
leaves the ferromagnetic state invariant  on a bipartite lattice. 
Thus, a level crossing  cannot be argued based on
identically the same symmetry argument. 
As it turns out, however, it is again possible to generalize 
the argument by using a more elaborate symmetry operator, which 
changes eigenvalue in going from the $U=0$ state to large $U$ state. 
In the square lattice this symmetry operator is easily identified.
It is the mirror symmetry across the diagonal of the square 
lattice $ L=l \times l$ 
with $l/2$ odd (see picture \ref{figsquare}).
In the honeycomb lattice one can also find a symmetry operation
with the same property. However, it
is much more involved and we will not discuss it in detail here. We 
only mention  that this symmetry is obtained by a $120^\circ$ rotation  
around a site followed by an additional gauge transformation 
$ c^{\dag}_i \to c^{\dag}_i e^{j \theta_i} $ with suitable angles 
$\theta_i$.\cite{parola}

Based on this analysis, we can therefore conclude that,
upon averaging over the boundary conditions, there will 
be, both in the square and in the honeycomb lattice, a level 
crossing when the system is in the Mott state, but none in the band
insulator state . Therefore, we should expect a polarization anomaly,
and a metallic state at the transition, also in these two dimensional cases.

However, before moving on to do numerical work and check these
expectations  in these more difficult problems, it is wise to solve them
in simple mean-field which, as the D=1 case demonstrated,\cite{ortiz2}
is always very instructive.

\section{2D bipartite lattices:  Hartree-Fock  approximation} \label{sec:hf}

We shall consider the Hubbard model on the bipartite  
honeycomb lattice, defined in  Sec. (\ref{sec:model}),  
and on the simpler square lattice. 
In the latter case, in order to remove the nesting degeneracy of the non 
interacting 2D  Fermi surface, we have also studied the effect of the  
next-nearest neighbor hopping.

In the Hartree -Fock approximation  the  ground state of the Hamiltonian 
is approximated by a Slater determinant. We may further 
assume that this Slater determinant $|SD>$  is 
factored in spin space 
\begin{equation}
|SD>= |SD_\uparrow> \otimes ~ |SD_\downarrow>
\end{equation}
With this choice it is simple to obtain the Hartree-Fock Hamiltonian 
by linearizing the interaction term:
$$U n_\uparrow n_\downarrow = U
 \left( <SD_\downarrow| n_\downarrow |SD_\downarrow>  n_\uparrow 
 + <SD_\uparrow| n_\uparrow |SD_\uparrow>  n_\downarrow \right) 
 - U <SD_\downarrow| n_\downarrow |SD_\downarrow>
 <SD_\uparrow| n_\uparrow |SD_\uparrow>
$$
In this way we obtain:
$$ H_{Hartree-Fock } = H_\uparrow + H_\downarrow + E_{const}
$$
where 
\begin{equation} \label{hfham}
H_\sigma = \sum_{k \in BZ} \epsilon_k c^{\dag}_{k B \sigma} c_{ k A \sigma}
+ {\rm h.c. } +\Delta_\sigma \sum_{k \in BZ} \left( c^{\dag}_{ k A \sigma } 
c_{k A \sigma }  - c^{\dag}_{ k B  \sigma} c_{k B \sigma}  \right)
\end{equation}
where we have employed a Fourier transform in the two sublattices 
and  correspondingly we have defined the complex function:   
\begin{equation} \label{defek}
\epsilon_k = \sum_{\mu} e^{ i \vec \xi_\mu \cdot \vec k} 
\end{equation}
In the square lattice case the presence of the next-nearest  hopping 
implies  a  further term: 
\begin{equation} \label{tprime}
H_{t^\prime}= \sum_{k \in BZ,\sigma  } 
\epsilon^\prime_k  (c^{\dag}_{k A \sigma}  c_{k A \sigma} 
+ c^{\dag}_{k B \sigma} c_{k B \sigma} )   
\end{equation} 
with $\epsilon^\prime_k=4 t^\prime \cos k_x \cos k_y$
  to be added  to $H_\sigma$ in Eq.(\ref{hfham}).
For a uniform solution in both sublattices the average spin densities 
are given by:
\begin{eqnarray}
<SD_\downarrow| n_{ \downarrow,R}  |SD_\downarrow> &=& 
\left\{ 
\begin{array}{c}   \rho_{ A \downarrow } ~~~ for~ R \in A  \\
                   \rho_{ B \downarrow } ~~~ for ~ R \in B 
\end{array}
\right.
                            \\
<SD_\uparrow| n_{ \uparrow,R}  |SD_\uparrow> &=& 
\left\{ 
\begin{array}{c}   \rho_{ A \uparrow } ~~~ for~ R \in A  \\
                   \rho_{ B \uparrow } ~~~ for ~ R \in B 
\end{array}
\right.
\end{eqnarray}
The parameters $\Delta_\uparrow$ and $\Delta_\downarrow$,  defining the 
Hartree -Fock hamiltonians are given therefore by:
\begin{eqnarray}
\Delta_\uparrow &=& {\Delta \over 2 } + { U \over 2 } 
( \rho_{A \uparrow }-
\rho_{ B \uparrow } ) \\
\Delta_\downarrow & =& { \Delta \over 2 } + { U \over 2 }
 ( \rho_{ A \downarrow}  -\rho_{ B \downarrow } ) 
\end{eqnarray}
The constant is obtained after little algebra 
$$ E_{const} = { U \over 2 } ( \rho_{A \uparrow } + \rho_{ B \uparrow} ) 
N_\downarrow + { U \over 2 } ( \rho_{A \downarrow } + \rho_{ B \downarrow} ) 
N_\uparrow  - { U L \over 2 } ( \rho_{ A \uparrow } \rho_{ A \downarrow } +
\rho_{ B \uparrow} \rho_{ B \downarrow} ) $$
where $N_\uparrow$ and $N_\downarrow $ are the total number of spin up and 
spin down particles  ($N_\uparrow=N_\downarrow=L/2$). 
At one electron per site all the bonding bands $E_{k,\sigma} = - \sqrt{ 
|\epsilon_k |^2 + \Delta_\sigma ^2 } $   are occupied 
 by the spin up and spin down electrons  and the total Hartree-Fock energy is
\begin{equation} 
E_{tot}  = E_{const} +    \sum\limits_{ k \in BZ \sigma } 
 E_{k,\sigma} 
\end{equation}
By minimizing the total energy 
we obtain two self consistent equations for the variational parameters 
$\Delta_\sigma$ :
\begin{equation} \label{selfhf}
 \Delta_\sigma = { \Delta  \over 2} - { \Delta_{-\sigma} \over L}
\sum_{k \in BZ }  { 1 \over E_{ k , -\sigma} } 
\end{equation}
These equations can be  easily solved by inserting a trial 
initial value for $\Delta_{\uparrow}$ and $\Delta_{\downarrow}$ 
in the rhs and iterating the result until selfconsistency is reached.
For small U only a self consistent solution with $\Delta_{\uparrow} = \Delta_{\downarrow}$
is possible, with a small 
 charge transfer from the 
electron-rich site $B$ 
to the electron-poor site $A$.
The plot of the quantity 
$\rho_A - \rho_B = (\Delta_{\uparrow} + \Delta_{\downarrow} - \Delta)^2/U$,
is shown in  Fig. (\ref{hfmag}).

For large $U$ a broken symmetry 
solution with $\Delta_\uparrow \ne \Delta_\downarrow $ is possible for 
$U$ sufficiently large. A finite value of the staggered magnetization 
$\sigma = { 1\over 2 } ( \rho_{A \uparrow } - \rho_{ A \downarrow } - \rho_{ B 
\uparrow} + \rho_{ B \downarrow} ) $ is stable, and is given by:
$$ \Delta_\downarrow - \Delta _\uparrow =  U \sigma $$
For large enough $U$, $\Delta$ can be neglected and we approach 
asymptotically the standard antiferromagnetic solution, where
$$ \Delta_\uparrow= - \Delta_\downarrow.$$
It is clear therefore that after the magnetic transition, 
which occurs first in HF,
there is a slightly larger  critical $U$ when one of the two bands 
becomes gapless with $\Delta_\sigma=0$. 
A plot of the Hartree-Fock band gaps  $\Delta _\uparrow$ 
and $ \Delta_\downarrow $ 
is shown  in Fig. (\ref{hfgap}).
The self consistent  equations (\ref{selfhf}) for the 2D 
square lattice case are not modified by 
the presence of $t^\prime$ at least in the insulating HF phases 
where the bonding bands 
$E^{B}_{k,\sigma}= \epsilon^\prime_k - E_{k,\sigma}$ 
are full and the antibonding bands 
$E^{A}_{k,\sigma}= \epsilon^\prime_k + E_{k,\sigma}$ 
are empty for both spin up and down electrons. 
However for $t^\prime \ne 0$   we found a finite $U$ interval --
between the band insulator at small $U$ and the Mott-Hubbard insulator
at large $U$ -- where the  bonding and 
the antibonding  bands do overlap in a finite energy range, leading
to a fully metallic behavior (see Fig.\ref{hfgaptp}). Here the 
above analysys should be slightly modified to take into account the gain 
in energy obtained by occupying some states of the antibonding band.

We will not study the square lattice tt' model further, since the 
general three-phase sequence: band insulator -- extended metal 
phase -- Mott-Hubbard insulator is not our main concern.
And although the Hartree-Fock approximation is by no means exact, it 
certainly represents a good starting point in dimension larger 
than one, so that we expect the tt' model to have indeed an extended metal
phase between the two insulators.   

According to Hartree-Fock, the two insulator phases are however
still adjacent on the two sides of $U_c$ in the square lattice 
with first neighbor hopping (t'=0), 
and in the honeycomb lattice, with any hopping. In the former, the 
nesting property of the non interacting Fermi surface affects 
dramatically the HF solution, making it effectively one dimensional, 
and leading to an unphysical first order transition.\cite{ortiz2} 
Moreover, even beyond Hartree-Fock, the first-neighbor square lattice model 
is not generic enough. An arbitrarily small next neighbor hopping
will take it away to a tt', where the two insulators are most likely
no longer adjacent.

We conclude that, at least at the HF level, the simplest model 
which generically possesses a transition between a Mott and a band 
insulator in 2D is the honeycomb lattice. Here
i) the HF transition to a magnetic phase is  second order; ii) 
there is no nesting of the non interacting Fermi surface, which consists
of just two k-points; and iii) 
the HF solution is always insulating (band or Mott)  as a function of $U$ 
with the excep[tion of a point where there is semimetallic behavior. 
For the above reasons, rather well justified at the HF level, in 2D 
we shall restrict our study to the honeycomb lattice case.

In the honeycomb HF solution, there is a small difference 
between the two critical values of $U$, one where magnetism sets in, and 
the next where a band gap closes, Fig. (\ref{hfgap}). Bearing in 
mind the discussion of Sec. [\ref{sec:chspgap}], we cannot judge 
whether this difference and its sign is 
just an artifact of the HF calculation  or a real one. 
Because of the crudeness 
of Hartree Fock, most likely the second.

The evolution of the full Hartree-Fock band structure for increasing
$U$ (not shown) is also instructive.
The bands  display a full gap
both in the band insulator, and in the magnetic insulator. The presence
of a spin gap in the latter is clearly an artifact due to breaking
of spin rotation invariance characteristic of Hartree-Fock. 
At the upper critical $U$, where the charge gap closes, there
is linear band dispersion, a point-like Fermi surface, and a 
Dirac massless spectrum centered at the 2D zone boundary. Therefore 
we could expect a semimetallic state to exist at the treshold between the 
two different insulators. 

\subsection{Nature of the 2D polarization anomaly in the 1D and 2D honeycomb 
lattice} \label{explanation}

We argue that close to the point where the effective gap
$\Delta_\downarrow$ changes sign, 
the polarization will change dramatically as a function of $U$.
To verify our statement, we make use of the theory of the polarization, 
in the geometric phase formulation \cite{ortiz1}, \cite{resta}, 
\cite{king}, \cite{vand}. In the Hartree-Fock case, this amounts to
calculate the contribution to the polarization of the spin-polarized band 
which becomes gapless at the critical point, when one effective 
gap parameter changes sign. 
In order to better illustrate that, we consider a model system 
consisting of spinless noninteracting fermions, described by the 
Hamiltonian (\ref{hfham})with a density of 1/2 fermion per site.
With periodic boundary conditions, we 
can calculate analytically the change of polarization as we vary 
continuously the fermion energy gap $\Delta$, or alternatively 
as we introduce a small dimerizing distortion $\delta$ in the hopping.
In the linear chain, without any distortion, $\delta \to 0$, the 
 variation of the polarization as a function of the 
energy difference  $\Delta$ between anions  and cations 
 is described by a   step function : $P_0(\Delta) = \frac{e a}{4}
\frac{\Delta}{|\Delta|}$ ( see Ref.\cite{ortiz1}).
When we vary $\Delta$ without changing its sign, the polarization 
does not change, while when the sign changes, the polarization
jumps by $\frac{e a}{2}$ as anticipated.
We introduce a dimerizing distortion (``optical phonon-like'') 
$\delta$ which  
amounts to a change of the hopping $t \to t(1\pm\delta)$ for
electrons hopping to the right and the left respectively of any $B$ 
site.
We can evaluate analytically the ``Born effective charge'', which is
the derivative of the polarization with respect to 
$\delta$ at $\delta =$ 0 :
\begin{equation}
\lim_{\delta \rightarrow 0} P'_{\delta}(\Delta) = \frac{e a}{2 \pi} \lbrace k' K - \frac{1}{k'}(2 K + E) \rbrace
\end{equation}
where $k = (1+{\Delta^2 \over 16 t^2})^{-1/2}$, $k' = {\Delta \over 4 t}( 1+{\Delta^2 \over 16 t^2}  )^{-1/2}$. $K = K(k)$ and $E = E(k)$ 
are complete elliptic functions of the first and second kind.

In this model, we have thus recovered the divergent effective charge 
anomaly discovered numerically in the 1D many-body calculation 
of Resta and Sorella \cite{sorella}.
The same analysys holds for two dimensional or higher d-dimensional lattices
provided that at the transition point $\Delta=0$ the  bonding and the
antibonding bands touch at a $d-1$ dimensional (Fermi-) surface,
(nesting). 
That is the case in particular for the $D=2$ square lattice 
but not for the honeycomb lattice.

In the honeycomb lattice, using the geometric phase 
technique for the mentioned spinless fermion model, it is straightforward 
to verify
that without any distortion the polarization does not change 
as we vary $\Delta$. 
Introducing a distortion along the $\xi_1$ direction: $t_0 =$ 
$t_2 =$ $t (1 - \delta)$, $t_1 =$ $t (1 + 2 \delta)$, 
one can then verify that the component of the change in polarization 
which is orthogonal to $\xi_1$ vanishes, at least in first order in
$\delta$. 

Along $\xi_1$, it is difficult to derive an analytic expression 
for the change in polarization .  We can however extract  
the singular part of the effective charge,   
in the neighbourhood of the zero of $\Delta$ 
\begin{equation}
\lim_{{\delta \rightarrow 0} \atop {\Delta \rightarrow 0}} P'_{\delta}(\Delta) = \frac{e a}{2 \pi} \frac{\Delta}{| \Delta |}
\end{equation}
The effective charge has therefore a finite symmetric jump 
when $\Delta$ changes sign. Since this is only the singular part, 
there will be in addition a smooth
background contribution shifting this jump with respect to zero.( In D=1,
the singularity was infinite, so there the shift was irrelevant).
 
\begin{figure} \caption{
honeycomb lattice, plot of the effective charge $Z^*$ as a function the Hubbard onsite interaction $U$ for $\Delta = 1$.
The effective charge shows a finite jump at $U_c$ unlike the 1D case where the jump diverged.
}
\label{z}
\end{figure}
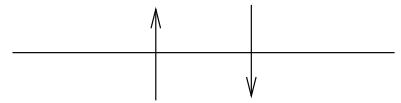
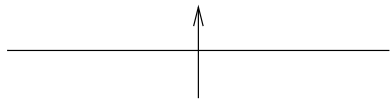
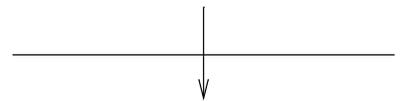

In conclusion, if the Hartree-Fock approximate picture were correct, 
we should expect such a polarization jump to emerge 
in the full many-body calculation for the 2D honeycomb lattice
of the next section. As it will turn out, the jump is indeed 
verified, see Fig. (\ref{z}). Moreover, the quantitative 
size of the jump is remarkably close to {\it twice} 
the value we have just obtained analytically. This result may be 
understood by  assuming that each of the two bands with opposite 
spins, are forced to close together their single particle gaps, due to
the spin rotation invariance of the model. This also suggests that the
 jump in the effective
 charge in 2D, similarly  to the jump in the polarization in 1D, maybe 
due to topological reasons and therefore rather general and weakly
dependent on the parameters defining the  model. 

\section{2D honeycomb lattice: many-body calculaton
 of the polarization} \label{sec:manybody}

We now proceed to do proper polarization--or more precisely, effective
charge-- calculation for a 2D 
interacting case. We restrict our study of the 
Hamiltonian  (\ref{hubbard}) to the honeycomb lattice ( a sort of 2D
hexagonal BN), 
which is slightly more interesting because it
does not possess inversion symmetry, and can have, {\it e.g.,}
a piezoelectric effect (see Sec.\ref{sec:discussion}).
The sites of the $A$ sublattice are given by 
$R_{A} = R_{m,n}$, with $R_{m,n} = (\frac{3}{2}n+3m,\frac{\sqrt{3}}{2}n)a$, 
$a$ is the lattice constant and $1\leq n \leq N_{y}$,  $1\leq m \leq \frac{N_{x}+2}{2}$.
The lattice is space periodic under translations 
$T_1=({3 \over 2}, {\sqrt{3} \over 2})N_y a$
and $T_2 = (3,0){N_x + 2 \over 2} a$. 
The reciprocal lattice vectors are $G_{1} = \frac{4 \pi}{\sqrt{3} a}(0,1)$ and
$G_{2} = \frac{2 \pi}{3 a}(1,-\sqrt{3})$. The sites of the $B$ 
sublattice are given by: $R_{B} = R_{mn}+\xi_{\mu}$, $\mu=0,1,2$
where the three vectors $\xi_{\mu}$ which connect neighouring 
$A$ and $B$ sites are $\xi_{0} = (- \frac{1}{2},\frac{\sqrt{3}}{2})a$, 
$\xi_{1} = (1,0)a$ and $\xi_{2} = (- \frac{1}{2},- \frac{\sqrt{3}}{2})a$. 
Obviously $\xi_{0}+\xi_{1}+\xi_{2} = 0$.
We can also consider, for the purpose of mimicking 
 a uniaxial stress along the $\xi_1$ direction: 
$t_0 = t_2 = t(1 - \delta)$ and $t_1 = t(1 + 2 \delta)$. 
The Hamiltonian depends parametrically on $\delta$, $H = H(\delta)$. 
In order to calculate the difference in the polarization of the system
as the parameter is varied between its values 0 and $\delta$, 
we consider families of Hamiltonians $H_k(\delta)$ obtained 
from (\ref{hubbard})
by introducing a complex hopping and substituting 
$t_{\mu} \rightarrow$ $t_{\mu} e^{i k \cdot \xi_\mu}$, 
where the parameter k is
given by $k=k_1 G_1 + k_2 G_2$, with $0 \le k_{\alpha} < 1$, ${\alpha} = 1,2$.
This is equivalent to imposing generalised boundary conditions 
on the original Hamiltonian 
\cite{gros}: If $\psi(r_1, \ldots, r_j, \ldots)$ is an
eigenfunction of $H$, then generalized boundary conditions imply 
$\psi(r_1, \ldots, r_j + T_{\alpha}, \ldots) = e^{i 2 \pi k_{\alpha}} \psi(r_1, \ldots, r_j, 
\ldots)$, ${\alpha} = 1,2$. Periodic boundary conditions 
correspond to  $k_{\alpha} = 0$ , antiperiodic to  $k_{\alpha} = 1/2$.
The polarization difference between two states which are 
characterised by the initial and final value of the distortion, 0 and $\delta$ 
is given as an integral over the generalised boundary conditions:
\begin{equation}
\Omega ~ G_{\alpha} \cdot \Delta P = e \int_0^1 d k_{\beta} \big( \Gamma_{\alpha}(\delta) - \Gamma_{\alpha}(0) \big)
\end{equation}
where $\Omega$ is the unit cell volume, $\alpha$ and $\beta$ take alternatively the values 1 or 2, $\Gamma_{\alpha}(\delta)$ 
is the many-body generalisation of the geometric phase:
\begin{equation}
\Gamma_{\alpha}(\delta) = i \int_0^1 d k_{\alpha} <\Phi_0(\delta, k)| \frac{\partial}{\partial k_{\alpha}} \Phi_0(\delta,k)>
\end{equation}
and $\Phi_0(\delta,k)$ is the ground state of $H_k(\delta)$, which satisfies periodic boundary conditions.

In the numerical calculation, we adopt a cell with 8 lattice 
sites ($N_x=N_y=2$) and 8 electrons. The number of spin-up and spin-down
electrons is 4, and the dimension of the Hilbert space is only 4900. 
The system is really very small, but we expect that averaging over 
the boundary conditions will reduce the
finite size effects.
In the calculation we are interested to study the behaviour of 
$\Delta P$, the polarization difference between 
the undistorted ($\delta=0$) and the distorted ($\delta \ne 0$) 
case, for different values of
the onsite interaction $U$.

We use a discretized form to calculate the geometric phase numerically
\begin{equation}
\Gamma_{\alpha}(\delta) = - i \ln \prod_{j=0}^{N_{\alpha} -1} 
\frac{ <\Phi_0(\delta, (\frac{\scriptstyle j}{\scriptstyle N_{\alpha}},k_{\beta}))|
 \Phi_0(\delta,(\frac{\scriptstyle j+1}{\scriptstyle  N_{\alpha}},k_{\beta}))> }
{| <\Phi_0(\delta, (\frac{\scriptstyle j}{\scriptstyle N_{\alpha}},k_{\beta}))|
 \Phi_0(\delta,(\frac{\scriptstyle j+1}{\scriptstyle  N_{\alpha}},k_{\beta}))>| }
\end{equation}
The geometric phase however is only defined modulo 2$\pi$ and 
an uncertainty in the result for the difference in polarization
arises. There will be an ambiguity in the value of the polarization 
difference up to a quantum. However the polarization difference 
should remain unambiguous for neighboring values of $\delta$. 
We can thus fix the 2$\pi$ uncertainty in $\Gamma(\delta)$ by requiring
that $\Gamma(\delta)$ is continuous in $\delta$ for a given value of $U$.
 We may also require that it is continuous for neighboring values of 
$U$ for fixed $\delta$, so  long as we do not cross a point of 
degeneracy in the electronic spectrum.\\
Convergence in the integration with respect to $k_{\beta}$ 
has not been easy and we had to use 
grids with 150 or 300 points to get reliable results.
At the end we found that in the undistorted case, 
similarly to the one-dimensional case, there is a critical 
point $U_{c}$, where the geometric 
phase changes discontinuously by $\pi$ for one $k_{\beta}$ in the 
2D Brillouin Zone. As in one dimension this 
effect maybe related to the presence  of  a level crossing 
 as a function of $U/t$ for a particular boundary 
condition.
Introducing next the dimerizing distortion, we found that 
the effective charge, as in the one-dimensional case, 
is positive in the region below $U_{c}$ and negative above. 
At $U_{c}$ it is
discontinuous, with a finite jump, instead of the one-dimensional divergence.
A plot of the effective charge as a function of $U/t$ 
 is shown in Fig. (\ref{z}).

\section{Piezoelectric Inversion, and Concluding Remarks}
\label{sec:discussion}

The transition between an ionic band insulator and a Mott-Hubbard
insulator as exhibited by the prototype model (Eq.\ref{hubbard}) is
quite interesting. We have extended existing studies in D= 1, probing
deeper into the transition. We have also 
carried out newer investigations in D=2, where same kind of
transition is shown to take place, with some differences between the
square and the honeycomb lattice.

With reference to the list of questions presented in the introduction,
we can now formulate the following answers.

1. The physical origin of the polarization anomaly is connected with
the symmetry switch between an even state, typical of the band insulator,
to an odd state, typical of the Mott insulator. The symmetry switch,
which we\cite{sorella} and others\cite{ortiz2} had noted earlier 
for finite size, is established here for arbitrary size. Moreover, 
in D=1, even and odd refer
to simple inversion symmetry, in D=2 the pertinent symmetry operation
is different and depends on the lattice.

2. The threshold state between the ionic and the Mott insulator, where the 
polarization abruptly changes sign, is one where the charge gap seems
to close, and is therefore metallic. This result underlines the 
profound difference between the two types of insulator, and invites
further studies, which are now beginning to appear\cite{fabrizio}
of this transition.

3. In D=2 the same model has again a band-to-Mott insulator transition.
The polarization anomaly is generally weaker, to a degree which depends on 
the lattice. In one dimension the effective charge diverges at 
the transition undergoing an abrupt sign change from $+\infty$ to $-\infty$.
In two dimensions this sign  change persists. We have shown  that 
in the 2D honeycomb lattice the divergence turns to a jump, again  with
a change of sign.
Here the magnitude of this jump  is found to coincide almost exactly with  
$ e a/\pi$, suggesting that this jump could be an experimentally 
detectable quantity dependent only on the bulk lattice constant and 
no other  details of the actual material.

4. The total piezoelectric coefficient $\gamma$ might be used to detect
the polarization anomaly, because it is sensitive to the sign of  $Z^*$ .
It is conventionally divided into two 
contributions:  the  first,  $\gamma_0$,  is purely electronic,
and is related  to  a uniform stress of the 
solid (a simple scaling of all distances  in the unit cell); 
the second, more important contribution, is obtained by keeping 
the unit cell fixed and by changing the distance between atoms.
In lattices without inversion symmetry, the piezoelectric coefficient
$\gamma$ is directly proportional, in magnitude {\em and sign},
to the effective 
charge $Z^*$, in the form (we omit here for simplicity 
tensorial and vectorial indices) \cite{martin,baroni}
\begin{equation} \label{piezo}
\bar \gamma = \gamma_0  + Z^* \xi
\end{equation}

where the  constant  $\xi$ represents the internal strain parameter.

An anomaly in $Z^*$ will clearly reflect in the piezoelectric coefficient.
Piezoelectric measurements could therefore
permit the detection of a band-to-Mott insulator transition, provided
the system does not possess inversion. 
The honeycomb lattice which we have studied in D=2 is the simplest 
one that does not possess inversion symmetry. Our study shows that 
the transition is not suppressed by piezoelectric strain.

We have not been able to identify, at present, a  likely
compound where this kind
of transition could be experimentally detected. It would seem possible,
nevertheless, that suitable systems could be engineered, especially
in the organic world.

\section{Acknowledgements}
We are thankful to M. Fabrizio, S. de Gironcoli,  A. Parola  and R. Resta
for useful discussions. N.G. acknowledges
financial support through an EEC
fellowship under the HCM Programme Contract No. ERBCHBGCT940721.
Work at SISSA was co-sponsored by INFM through PRA HTSC,
and by the European Union through Contract ERBCHRXCT940438.
Part of the calculations were performed during N.G.'s  visit 
to CINECA, under the Icarus2 project, Contract No.  ERBFMGECT950052.
 Stefano Cozzini is acknowledged for his help with the parallelization of the
 codes.

\begin{figure} \caption{ The model at the classical level when the hopping $t$
 is set to zero. The energy difference between an 
$A$ site and a $B$ site is $\Delta$. At 1 electron per site filling, when the onsite interaction $U<\Delta$ the
$B$ site is doubly occupied and the $A$ site empty. When $U>\Delta$ both sites are single occupied. 
}
\label{class}
\end{figure}

\begin{figure} \caption{
Calculation on 12 site rings (squares) and 6 site rings (triangles) of the
charge (black) and spin (white) gaps, using closed shells at fillings of 
1 electron per site. The data of the gaps for the 8, 10 and 12 sites are used 
 for a finite size scaling extrapolation to the infinite number of sites. 
The curves of the charge (solid line) and spin (dashed line) gaps are shown. 
The two curves seem to coincide in the region below $U_c$
}
\label{spinchargecl}
\end{figure}

\begin{figure} \caption{
Calculation on 12 site rings (solid and long dashed line) and 6 site rings 
(short dashed and dotted lines) of the
charge (long dashed and dotted lines) and spin (solid and short dashed lines) 
gaps, using open shells at fillings of 1 electron
per site. The results of the 12 site calculation practically coincide with the 
finite size scaling results of the closed shells.
}
\label{spinchargeop}
\end{figure}

\begin{figure} \caption{
Square lattice of side $l$ with $l/2$ odd. The number of lattice sites which
 lie on either side of the diagonal is odd.
}
\label{figsquare}
\end{figure}

\begin{figure} \caption{
honeycomb lattice, Hartree-Fock calculation of the staggered magnetization $\sigma$ and the density difference $\rho_A-\rho_B$.
The transition is 2nd order. The arrow indicates the transition point. 
}
\label{hfmag}
\end{figure}

\begin{figure} \caption{
honeycomb lattice, Hartree-Fock calculation of the effective energy gap $\Delta_{\sigma}$.
Two different critical values $U_{c1}$, $U_{c2}$ of the interaction are identified. 
Below $U_{c1}$ there is only a solution with $\Delta_{\uparrow}=\Delta_{\downarrow}$.
Above $U_{c1}$ a solution with $\Delta_{\uparrow} \ne \Delta_{\downarrow}$ exists leading to a 
finite value of the magnetization $\sigma$. One of the spin bands becomes
gapless at a slightly different value of the interaction $U_{c2}$.
}
\label{hfgap}
\end{figure}

\begin{figure} \caption{Square lattice with nearest neighbor hopping:
HF calculation of  the staggered magnetization $\sigma$,  
the density difference $\rho_A-\rho_B$  and the band insulator gap 
$E_G$}
\label{hfgaptp} 
\end{figure}

\end{document}